\documentclass{moriond}

\usepackage{wrapfig}
\def\Journal#1#2#3#4{{#1} {\bf #2}, #3 (#4)}

\def\NIMA{{\em Nucl. Instrum. Methods} A}

\def\PLB{{\em Phys. Lett.}  B}
\def\PRL{\em Phys. Rev. Lett.}
\def\PRD{{\em Phys. Rev.} D}

\def\JHEP{{\em JHEP}}
\def\JINST{{\em JINST}}
\def\JCAP{{\em JCAP}}

\def\ber{$^7$Be\ }
\def\bor{$^8$B\ }
\def\cro{$^{51}$Cr\ }
\def\cer{$^{144}$Ce-$^{144}$Pr\ }
\newcommand{\Fluka}{F{\sc{luka}} }

\newcommand{\Geant}{G{\sc{eant}}}

\newcommand{\Photo}{}

\begin{document}
\title{Recent Borexino results and prospects for the near future}

\author{ 
{ \small
D.~D{\textquoteright}Angelo$^h$,
G.~Bellini$^h$,
J.~Benziger$^k$,
D.~Bick$^s$,
G.~Bonfini$^e$,
M.~Buizza Avanzini$^h$,
B.~Caccianiga$^h$,
L.~Cadonati$^p$,
F.~Calaprice$^l$,
P.~Cavalcante$^e$,
A.~Chavarria$^l$,
A.~Chepurnov$^r$,
S.~Davini$^c$,
A.~Derbin$^m$,
A.~Empl$^t$,
A.~Etenko$^g$, 
F.~von Feilitzsch$^n$,
K.~Fomenko$^{b,e}$,
D.~Franco$^{a}$,
C.~Galbiati$^l$,
S.~Gazzana$^e$,
C.~Ghiano$^e$,
M.~Giammarchi$^h$,
M.~G\"{o}ger-Neff$^n$,
A.~Goretti$^l$,
L.~Grandi$^l$,
C.~Hagner$^s$,
E.~Hungerford$^t$,
Aldo Ianni$^e$,
Andrea Ianni$^l$,
V.~Kobychev$^f$,
D.~Korablev$^b$,
G.~Korga$^t$, 
D.~Kryn$^a$,
M.~Laubenstein$^e$,
B.~Lehnert$^w$,
T.~Lewke$^n$,
E.~Litvinovich$^{g,v}$,
F.~Lombardi$^e$,
P.~Lombardi$^h$,
L.~Ludhova$^h$,
G.~Lukyanchenko$^g$,
I.~Machulin$^{g,v}$,
S.~Manecki$^q$,
W.~Maneschg$^i$,
G.~Manuzio$^c$,
Q.~Meindl$^n$,
E.~Meroni$^h$,
L.~Miramonti$^h$,
M.~Misiaszek$^d$,
P.~Mosteiro$^l$,
V.~Muratova$^m$,
L.~Oberauer$^n$,
M.~Obolensky$^a$,
F.~Ortica$^j$,
K.~Otis$^p$,
M.~Pallavicini$^c$,
L.~Papp$^{e,q}$,
L.~Perasso$^c$,
S.~Perasso$^c$,
A.~Pocar$^p$,
G.~Ranucci$^h$,
A.~Razeto$^e$,
A.~Re$^h$,
A.~Romani$^j$,
N.~Rossi$^e$,
R.~Saldanha$^l$,
C.~Salvo$^c$,
S.~Sch\"onert$^{n}$,
H.~Simgen$^i$,
M.~Skorokhvatov$^{g,v}$,
O.~Smirnov$^b$,
A.~Sotnikov$^b$,
S.~Sukhotin$^g$, 
Y.~Suvorov$^{u,g}$,
R.~Tartaglia$^e$,
G.~Testera$^c$,
D.~Vignaud$^a$,
R.B.~Vogelaar$^q$,
J.~Winter$^n$,
M.~Wojcik$^d$,
A.~Wright$^l$,
M.~Wurm$^s$,
J.~Xu$^l$,
O.~Zaimidoroga$^b$,
S.~Zavatarelli$^c$,
K.~Zuber$^{w}$,
G.~Zuzel$^{d}$
}
}

\address{{\tiny
a) Laboratoire AstroParticule et Cosmologie, 75231 Paris cedex 13, France\\
b) Joint Institute for Nuclear Research, Dubna 141980, Russia\\
c) Dipartimento di Fisica, Universit\`{a} e INFN, Genova 16146, Italy\\
d) M. Smoluchowski Institute of Physics, Jagellonian University, Krakow, 30059, Poland\\
e) INFN Laboratori Nazionali del Gran Sasso, Assergi 67010, Italy\\
f) Kiev Institute for Nuclear Research, Kiev 06380, Ukraine\\
g) NRC Kurchatov Institute, Moscow 123182, Russia\\
h) Dipartimento di Fisica, Universit\`{a} degli Studi e INFN, Milano 20133, Italy\\
i) Max-Plank-Institut f\"{u}r Kernphysik, Heidelberg 69029, Germany\\
j) Dipartimento di Chimica, Universit\`{a} e INFN, Perugia 06123, Italy\\
k) Chemical Engineering Department, Princeton University, Princeton, NJ 08544, USA\\
l) Physics Department, Princeton University, Princeton, NJ 08544, USA\\
m) St. Petersburg Nuclear Physics Institute, Gatchina 188350, Russia\\
n) Physik Department, Technische Universit\"{a}t M\"{u}nchen, Garching 85747, Germany\\
p) Physics Department, University of Massachusetts, Amherst MA 01003, USA\\
q) Physics Department, Virginia Polytechnic Institute and State University, Blacksburg, VA 24061, USA\\
r) Institute of Nuclear Physics, Lomonosov Moscow State University, 119899, Moscow, Russia\\
s) Institut f\"ur Experimentalphysik, Universit\"at Hamburg, Germany\\
t) Department of Physics, University of Houston, Houston,  TX 77204, USA\\
u) Physics and Astronomy Department, University of California Los Angeles (UCLA), Los Angeles, CA 90095, USA\\
v) National Research Nuclear University MEPhI (Moscow Engineering PhysicsInstitute), 31 Kashirskoe Shosse, Moscow, Russia\\
w) Institut f\"ur Kern- und Teilchenphysik, Technische Universit\"at Dresden, Dresden 01069, Germany\\
$\sim$
}}

\maketitle

\abstracts{
The Borexino experiment, located in the Gran Sasso National Laboratory, is an organic liquid scintillator detector conceived for the real time spectroscopy of low energy solar neutrinos. The data taking campaign phase I (2007 - 2010) has allowed the first independent measurements of \ber, \bor and pep fluxes as well as the first measurement of anti-neutrinos from the earth. After a purification of the scintillator, Borexino is now in phase II since 2011. We review here the recent results achieved during 2013, concerning the seasonal modulation in the \ber signal, the study of cosmogenic backgrounds and the updated measurement of geo-neutrinos. We also review the upcoming measurements from phase II data (pp, pep, CNO) and the project SOX devoted to the study of sterile neutrinos via the use of a \cro neutrino source and a \cer antineutrino source placed in close proximity of the active material.}

\section{Why solar neutrinos with Borexino}
\label{sec:intro}

\begin{figure}
\begin{minipage}{0.49\linewidth}
\centerline{\includegraphics[width=\linewidth]{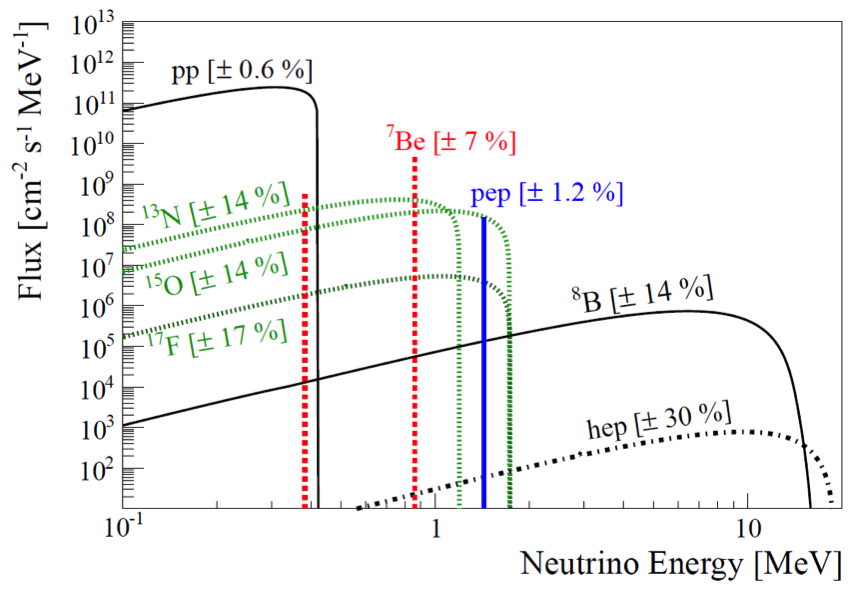}}
\caption{Solar Neutrino Spectrum.}
\label{fig:spectrum}
\end{minipage}
\hfill
\begin{minipage}{0.49\linewidth}
\centerline{\includegraphics[width=\linewidth]{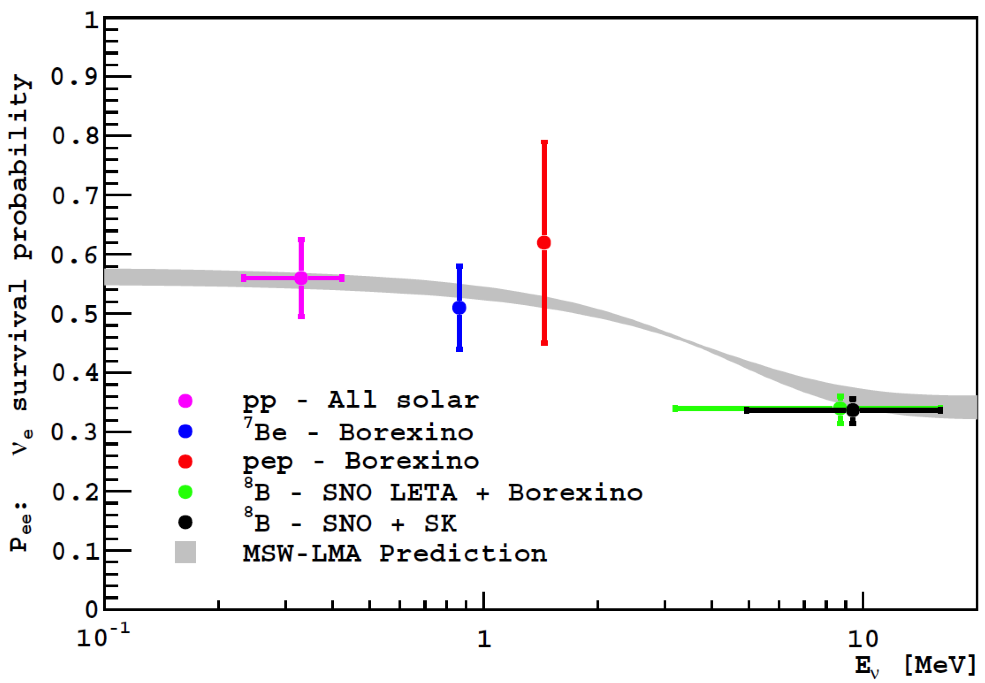}}
\caption{Survival probability of electron neutrinos as foreseen by the MSW-LMA model. Data points discussed here are also shown.}
\label{fig:pee}
\end{minipage}
\end{figure}

The Sun is an intense source of neutrinos, produced in nuclear reactions of the p-p chain and of the CNO cycle \cite{hax}.
Measurements of the individual neutrino fluxes is of paramount importance for both particle physics and astrophysics.
The solar neutrino spectrum can be seen in fig. \ref{fig:spectrum}. 
Up to a few years ago, spectroscopical measurements were performed by water Cherenkov detectors above $\sim$5Mev 
and concerned only \bor neutrinos for less then 1\% of the total flux. 
The bulk of neutrinos at low energies were detected with radiochemical experiments, incapable of resolving the individual components.
Neutrinos are emitted in the sun as electron flavour neutrinos and oscillate to a different flavour during the trajectory to the Earth. 
The MSW mechanism at Large Mixing Angle (LMA) foresees the survival probability for electron neutrinos on Earth (fig. \ref{fig:pee}). 
Borexino was designed to achieve spectroscopy of the low energy part of the solar neutrino spectrum, in particular the flux of the \ber monochromatic line at 862keV. 
Borexino has largely exceeded the expected performance with the physics program broadening way past the original goal.

\section{The Borexino Project}
\label{sec:detector}

\begin{wrapfigure}{r}{0.5\textwidth}
\begin{center}
\includegraphics[width=\linewidth]{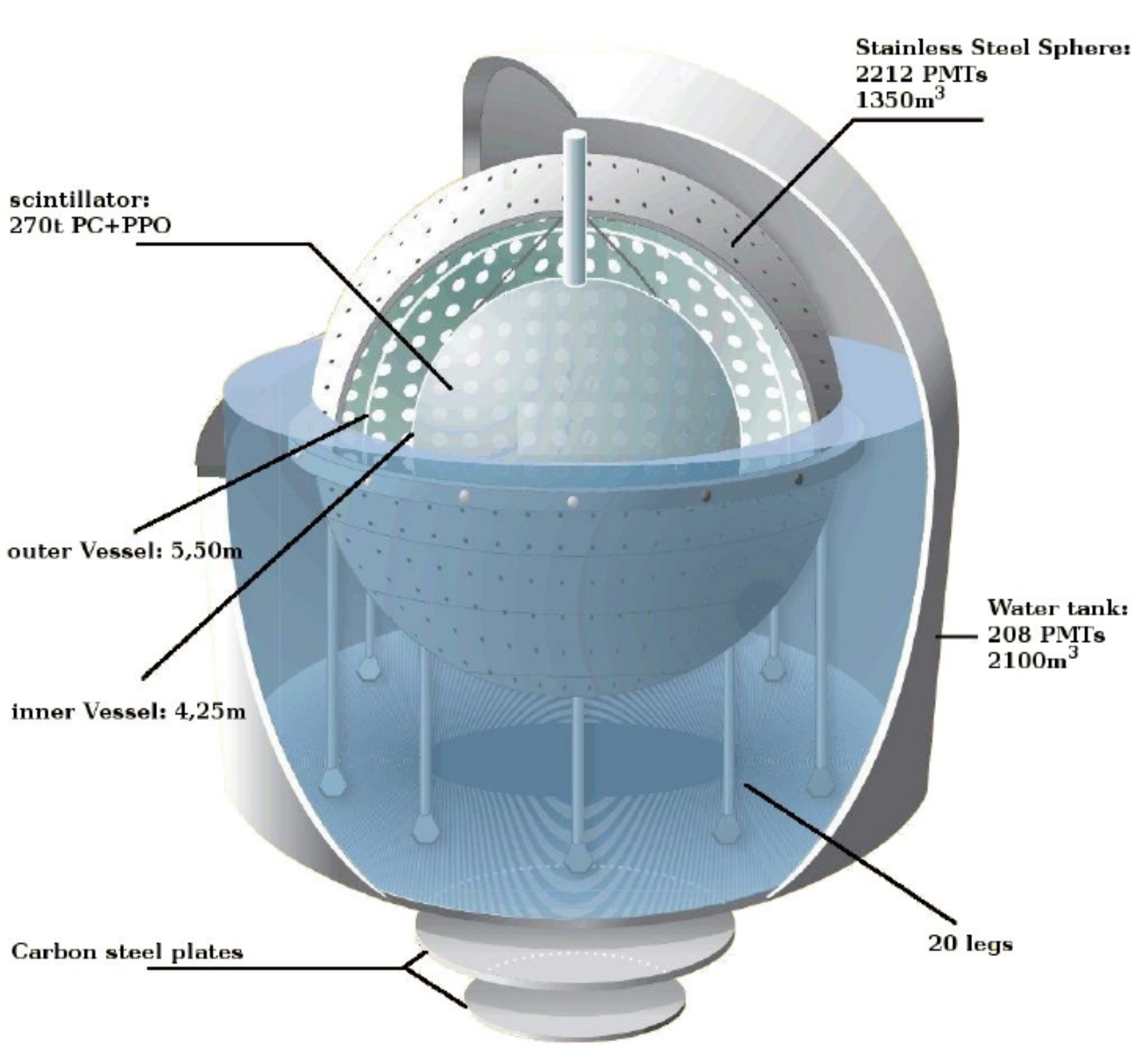}
\caption{Schematics of the Borexino detector.}
\label{fig:detector}
\end{center}
\end{wrapfigure}

\paragraph{Detector layout.}
The Borexino detector \cite{bx} is sketched in fig. \ref{fig:detector}. 
It is located at the Gran Sasso National Laboratories (LNGS) in central Italy, at a depth of 3800m w.e..
The active mass consists of 278t of pseudocumene (PC), doped with 1.5g/l of PPO. 
The scintillator is contained in a thin (125 $\mu$m) nylon Inner Vessel (IV), 8.5m in diameter.
The IV is surrounded by two concentric PC buffers doped with a light quencher. 
The scintillator and buffers are contained in a Stainless Steel Sphere (SSS) with a diameter of 13.7m. 
The SSS is enclosed in a Water Tank (WT), containing 2100t of ultra-pure water as an additional shield. 
The scintillation light is detected via 2212 8" PhotoMultiplier Tubes (PMTs) uniformly distributed on the inner surface of the SSS. 
Additional 208 8" PMTs instrument the WT and detect the Cherenkov light radiated by muons in the water shield. 
Neutrinos are detected via elastic scattering on electrons in the target material.

\paragraph{Detector performance.}
The Borexino scintillator has a high light yield: $\sim$10$^4$ photons/MeV, resulting in 500 detected photoelectrons/MeV.
The fast time response ($\sim$ 3ns) allows to reconstruct the events position by means of a time-of-flight technique with $\sim$13cm precision.
Depending on the analysis, the fiducial volume is defined between 75t and 150t.
The signature of \ber neutrinos is a Compton-like shoulder at 665keV in the electron recoil spectrum.
The energy resolution (1$\sigma$) at \ber energy is as low as 44 keV (or 6.6\%).
The energy threshold is $\sim$40keV, however the analysis threshold was limited so far by the knowledge of $^{14}$C spectral shape to $\sim$250keV.
Recently we have successfully reduced our threshold to 165keV (sec. \ref{sec:phaseII}).
Pulse Shape Analysis (PSA) is performed to identify various classes of events, 
among which electronic noise, pile-up events, muons, $\alpha$ and $\beta$ particles. 

\paragraph{Borexino timeline.}
\label{sec:history}

The Phase I of Borexino occurred between May 2007 and May 2010. 
After the purification of the scintillator through loop water extraction, 
in November 2011 the Phase II of Borexino started and is expected to last at least until the end of 2014.
The energy scale uncertainty in the range [0.2-2.0]MeV has been determined at 1.5\% precision, using multiple gamma sources. 
The position reconstruction algorithm has been tuned using a Rn source placed in 184 positions inside the active volume. 
The determination of the fiducial volume is -1.3\%+0.5\%. 
An external calibration campaign has allowed to correctly model external gamma backgrounds. 
The calibrations will be repeated before the end of Phase II to demonstrate stability of the detector and possibly to improve the previous results\cite{calib}.

\begin{table}
\begin{center}
\begin{tabular}{ccccc}
\hline
Isotope&Typical&Required&Before purification&After purification\\
\hline
$^{238}$U               &2 10$^{-5}$g/g (dust)         &$\leq$10$^{-16}$g/g&(5.3$\pm$0.5) 10$^{-18}$ g/g     &$<$ 0.8 10$^{-19}$g/g\\
$^{232}$Th              &2 10$^{-5}$g/g (dust)        &$\leq$10$^{-16}$g/g&(3.8$\pm$0.8) 10$^{-18}$ g/g     &$<$ 1.0 10$^{-18}$g/g\\
$^{14}$C/$^{12}$C&10$^{-12}$ (cosmogenic)&$\leq$10$^{-18}$      &(2.69$\pm$0.06) 10$^{-18}$       &unchanged\\
$^{222}$Rn             &100 atoms/cm$^3$ (air)    &$\leq$10cpd/100t     &~1cpd/100t                                      &unchanged\\
$^{40}$K                  &2 10$^{-6}$g/g (dust)         &$\leq$10$^{-18}$ g/g&$\leq$0.4 10$^{-18}$ g/g            &unchanged\\
$^{85}$Kr                 &1Bq/m$^3$ (air)                 &$\leq$1cpd/100t        &(30$\pm$5) 10cpd/100t                &$\leq$5cpd/100t\\
$^{39}$Ar                 &17mBq/m$^3$ (air)           &$\leq$1cpd/100t         &$\ll^{85}$Kr                                    &$\ll^{85}$Kr\\
$^{210}$Po             &                                              &not specified                &$\sim$80 $\rightarrow$ $\sim$5cpd/t                        &unchanged\\
$^{210}$Bi              &                                              &not specified                &$\sim$10 $\rightarrow$ $\sim$75cpd/100t               &(27$\pm$8)cpd/100t\\
\hline
\end{tabular}
\caption{Background levels in the Borexino scintillator. cpd stands for "counts per day".}
\label{tab:backgrounds}
\end{center}
\end{table}

\paragraph{Borexino backgrounds.}
\label{sec:backgrounds}
The background levels of the Borexino scintillator are summarised in tab. \ref{tab:backgrounds}. 
$^{238}$U and $^{232}$Th contaminations are unprecedented and largely exceed the requirements of the experiment. 
Most other contaminants are at acceptable levels or have been reduced below required limits after purification. 
We observe the presence of $^{210}$Bi and $^{210}$Po out of equilibrium. 
$^{210}$Po is of little concern as it is an $\alpha$ emitter and it can be identified by PSA. 
$^{210}$Bi has been rising during the data taking due to unclear motivations, possibly related to the movement of the scintillator. 
This variation was modelled and taken into account during the \ber flux modulation analysis (sec. \ref{sec:annual}). 
$^{210}$B was later reduced by the purification campaign to a level which might allow the measurement of CNO neutrino flux.

\section{Early results}

\paragraph{\ber flux.}
\label{sec:ber}

\begin{wrapfigure}{r}{0.5\textwidth}
\vspace{-0.5cm}
\begin{center}
\includegraphics[width=\linewidth]{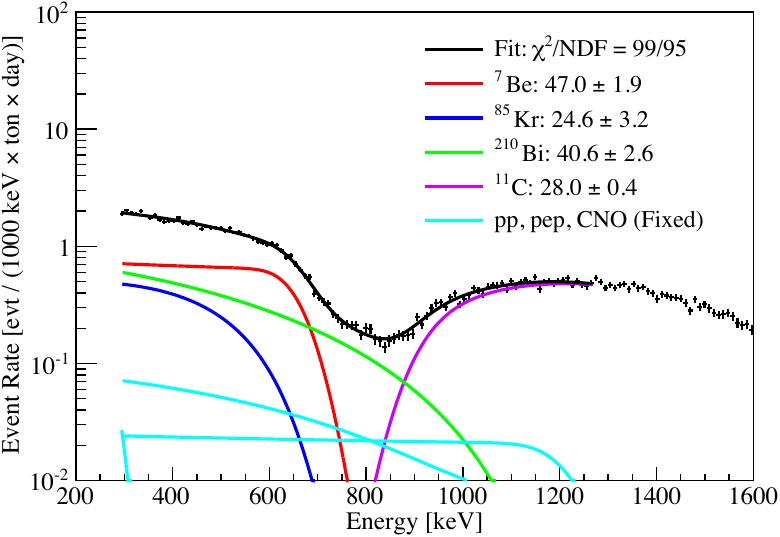}
\caption{Fit of the energy spectrum for \ber flux analysis.}
\label{fig:be7}
\end{center}
\end{wrapfigure}

The measurement of the flux of \ber neutrinos was the primary goal of Borexino. 
Thanks to the unprecedented background levels, the first observation was published in summer 2007, after only 3 months of data. 
The flux was later reviewed in 2008 and 2011 reducing the error every time.
The signal is extracted by a spectral fit along with other neutrino signals and background components (fig. \ref{fig:be7}).
The fit is performed both using analytical spectral shapes and MonteCarlo generated curves, 
with or without the statistical subtraction of $\alpha$ events via PSA, obtaining the same result within errors.
The last measured rate\cite{ber} after 741d of live time (full Phase I) is $R_{Be}=46.0\pm1.5_{stat}\pm1.6_{syst}$/d/100t. 
For the first time the experimental error (4.8\%) is lower the theoretical error (7\%). 
The rate corresponds to a flux of $\Phi_{Be} = (3.10\pm0.15) \times 10^9$cm$^{-2}$s$^{-1}$ with a survival probability $P_{ee} = 0.51\pm0.07$ at 862keV.

\paragraph{\ber day-night asymmetry.}
\label{sec:daynight}

We have carefully inspected the data set at the energy of \ber neutrino scattering looking for an eventual day-night asymmetry\cite{day}.
This was foreseen in an alternative MSW scenario, called LOW, also compatible to some extent with the solar neutrino results. 
The LOW scenario before Borexino could be excluded only assuming CPT invariance and using the KamLAND reactor's antineutrino results.
Our data is consistent with no asymmetry: $A_{DN} = {\Phi_N - \Phi_D} / \Phi = 0.001\pm0.012_{stat}\pm0.007_{syst}$.
With this result, the MSW-LOW mechanism can be ruled out at 8.5$\sigma$ using only solar neutrino data.

\paragraph{\bor flux}
\label{sec:bor}

Borexino has measured the \bor flux down to 3.0MeV\cite{bor}. 
While the much larger water Cherenkov detectors can achieve better precision, Borexino holds the lowest threshold achieved so far. 
Lowering the threshold on \bor spectroscopy represents one of the key features to inspect the transition region of the LMA solution (see below).
The measured rate is $R_{B} = 0.22\pm0.04_{stat}\pm0.01_{syst}$ and corresponds to a flux of $\Phi_{B}=(2.4\pm0.4_{stat}\pm0.1_{syst}) \times 10^6$cm$^{-2}$s$^{-1}$. The flux above 5MeV is in good consistency with other results. 

\paragraph{pep flux and CNO limits.}
\label{sec:pep}

As shown in fig. \ref{fig:pee}, the pep neutrino energy lies at the boundary between the Vacuum and the Transition region of the MSW survival probability.
Pep neutrinos are closely related to the fundamental pp neutrinos and have their flux theoretically well constrained by this relation. 
Measuring the pep neutrino flux therefore can also test the core of the Standard Solar Model.
In the same energy region are neutrinos from the CNO cycle reactions. 
CNO neutrinos are poorly constrained by the SSM and have never been detected so far. 
At this energy the cosmogenic background of $^{11}$C is overwhelming the pep/CNO neutrino flux by about an order of magnitude. 
We have made the measurement\cite{pep} possible exploiting the three-fold coincidence between the parent muon, the $^{11}$C and the neutron most often accompanying its production to suppress the background.
The rate of pep neutrinos has been extracted with a multivariate analysis based of the energy of the event, the distance from the center of the detector and a pulse shape parameter\cite{pep}. 
The rate is $R_{pep}=(3.1\pm0.6_{stat}\pm0.3_{syst})$ /d/100t corresponding to a flux of $\Phi_{pep}=(1.6\pm0.3) \times 10^8$cm$^{-2}$s$^{-1}$ and a survival probability of $P^{ee}_{pep} = 0.62\pm0.17$ at 1.44MeV.
The flux of CNO neutrinos could not be extracted due to the spectral shape degeneracy with the $^{210}$Bi background. 
The strongest upper limit available to date has been however obtained from this analysis. 
The rate of CNO neutrinos is $R<7.1 /$d/100t at 95\% C.L. corresponding to $\Phi_{CNO} < 7.7 \times 10^8$cm$^{-2}$s$^{-1}$.

\paragraph{Survival probability after Borexino.}
\label{sec:pee}

Fig. \ref{fig:pee} shows the survival probability of electron neutrinos emitted by the Sun after travelling to the Earth along with experimental data available after the Borexino measurements. 
In addition to the measurements already discussed, the fundamental pp flux has been better determined (value shifted and errors reduced) by subtracting from the integrated measured rate of radiochemical experiments the other signal components, in particular \ber as measured by Borexino. 
The unexplored transition region 1-3MeV still has room for alternative models and new physics, in particular the proposed Non Standard neutrino Interactions (NSI) which foresee a different transition shape. 
There are two ways to test these hypotheses or confirm LMA: reducing the errors on pep flux (and to a lower extent on \ber flux) and lowering the threshold on \bor neutrinos to observe (or not) the expected upturn of the spectrum. 
In its Phase II Borexino will follow both approaches.

%
\newpage

\section{Results in 2013}
\paragraph{\ber flux annual modulation.}
\label{sec:annual}

\begin{figure}
\begin{center}
\includegraphics[width=0.49\linewidth]{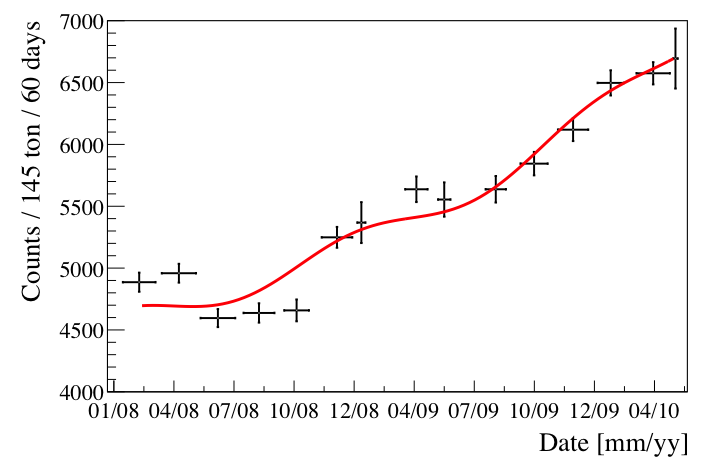}
\includegraphics[width=0.50\linewidth]{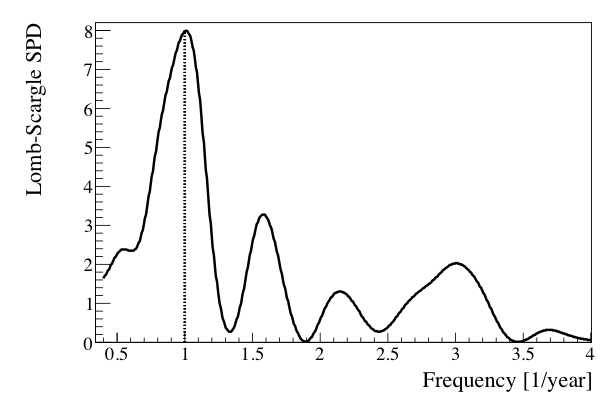}
\caption{\ber flux modulation analysis. Left: 60d binned count rate evolution. Right: 10d binned Lomb-Scargle Spectral Power Density.} 
\label{fig:annual}
\end{center}
\end{figure}

The solar neutrino flux is expected to undergo a yearly modulation due to the eccentricity of the Earth's orbit around the Sun. 
The flux is minimal at the beginning of July and maximal at the beginning of January. The expected amplitude is $\pm$3.4\%.
The observation of this modulation in the \ber flux is the ultimate proof that Borexino is actually observing neutrinos from the Sun.
For this analysis\cite{long} we have defined a dynamic and enlarged 141t FV with respect to the 75t used in the \ber rate analysis. 
This has been possible thanks to the precise determination of the vessel's shape from the distribution of $^{210}$Bi events deposited on the vessel's surface on a weekly basis.
Nevertheless performing the spectral fit on sub-periods of the data set is not a viable method due to reduced statistics.
Using the Phase I data set (850d astr. time), we have considered the count rate in the \ber region (105-380p.e.), 
which also includes background from the decay of $^{210}$Bi in the scintillator.
The last has been rising exponentially during the Phase I (sec \ref{sec:backgrounds}).
We have averaged the rate on a 60d base (fig. \ref{fig:annual}, left) and we have fitted the result with:
\begin{equation}
R=R_0+R_{Bi}e^{\Lambda_{Bi} t} + \overline{R}[1+2\epsilon \cos (\frac{2\pi t}{T}-\phi)]
\end{equation}
where $R_{0}$ accounts for a time independent background component.
The rate $R_{Bi}$ and the time constant $\Lambda_{Bi}$ of $^{210}Bi$ background are fixed to values independently determined from a different energy interval. 
The period found is $T=(1.01\pm0.07)$y and the phase, measured from Jan 1$^{st}$ 2008, is $\phi=(11.0\pm4.0)$d. 
The average \ber rate and the eccentricity $\epsilon$ are consistent within 2$\sigma$ with the spectral fit result and with the expected orbit eccentricity, respectively. 
The hypothesis  of no modulation is rejected at $>3\sigma$.
An alternative approach uses Lomb-Scargle frequency analysis in 10d binned data set. 
The Spectral Power Density (SPD) distribution is shown in fig. \ref{fig:annual} (right). 
The significance of SPD peaks has been evaluated by Monte Carlo simulations of the signal and the background. 
The peak at 1y with SPD=7.96 has significance largely above $3\sigma$, while no other peak exceeds $2\sigma$.

\paragraph{Cosmogenics}
\label{sec:cosmo}

\begin{wrapfigure}{r}{0.5\textwidth}
\begin{center}
\includegraphics[width=\linewidth]{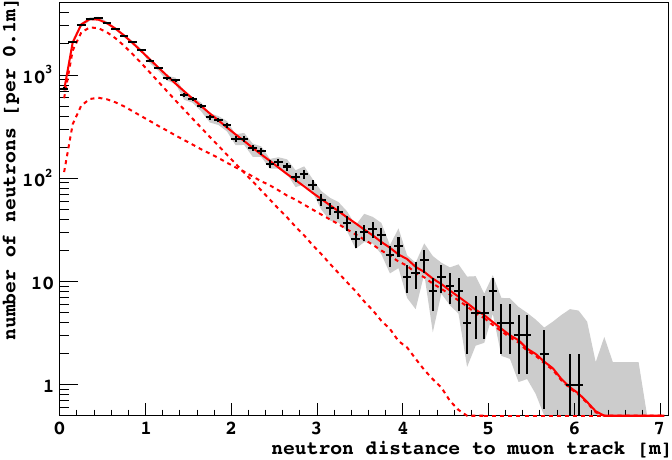}
\caption{Lateral distance of neutron capture points from the parent muon track. Details of the fit can be found in ref. 9.}
\label{fig:lateral}
\end{center}
\end{wrapfigure}

We have performed a thorough study of cosmogenic backgrounds in Borexino\cite{cosmo}.  
The results are not only essential to low-energy neutrino analyses, but are also of  substantial interest for direct dark matter and $0\nu\beta\beta$ searches at underground facilities. 
Based on thermal neutron captures
in the scintillator, a spallation neutron yield of
$Y_{n} = (3.10 \pm 0.11) \cdot 10^{-4}\,n/(\mu \cdot (\rm{g/cm}^{2}))$ was
determined. The lateral distance profile was measured based on
the reconstructed parent muon tracks and neutron capture vertices and is shown in fig. \ref{fig:lateral}.
An average lateral distance of 
$\lambda=(81.5\pm2.7)$\,cm was found.

\noindent The data results on neutron yield, multiplicity and lateral distributions were compared to
Monte Carlo simulation predictions by the \Fluka and \Geant4 frameworks
and are largely compatible. The simulated neutron
yield of \Fluka shows a deficit of $\sim$20\,\%, 
while the result of the \Geant4 simulation is in  good agreement with
the measured value. However, both simulations should be increased as a result 
on an underprediction of $^{11}$C production.
The production rates of several cosmogenic radioisotopes in the
scintillator were determined based on a simultaneous fit to
energy and decay time distributions. Results of a corresponding
analysis performed by the KamLAND collaboration are similar to our findings.
Moreover, Borexino rates were compared to predictions by \Fluka and \Geant4: While there is good
agreement within their uncertainties for most 
isotopes, some cases ({$^{12}$B}, {$^{11}$C}, {$^{8}$Li} for both
codes and {$^{8}$B}, {$^{9}$Li} for \Geant4 only) show a significant
deviation between data and Monte Carlo simulation predictions.  

\paragraph{Geo-neutrinos}
\label{sec:geo}

\begin{figure}[b!]
\begin{center}
\includegraphics[width=0.53\linewidth]{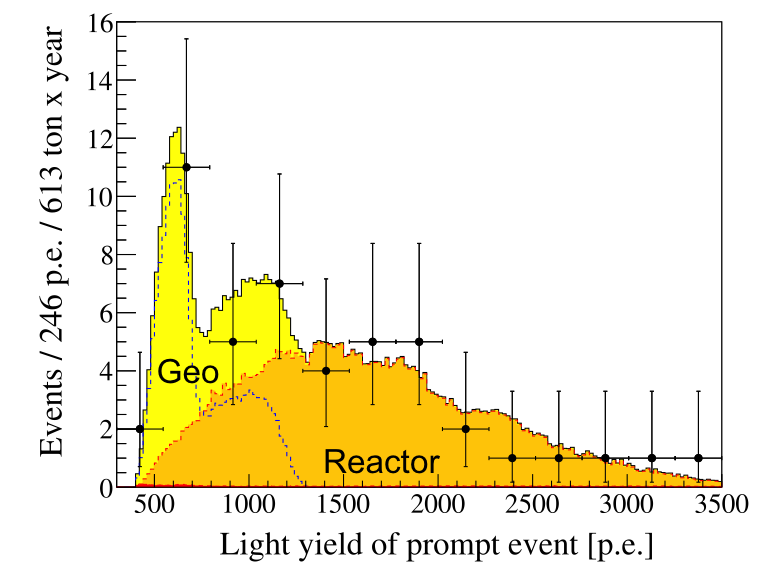}
\includegraphics[width=0.46\linewidth]{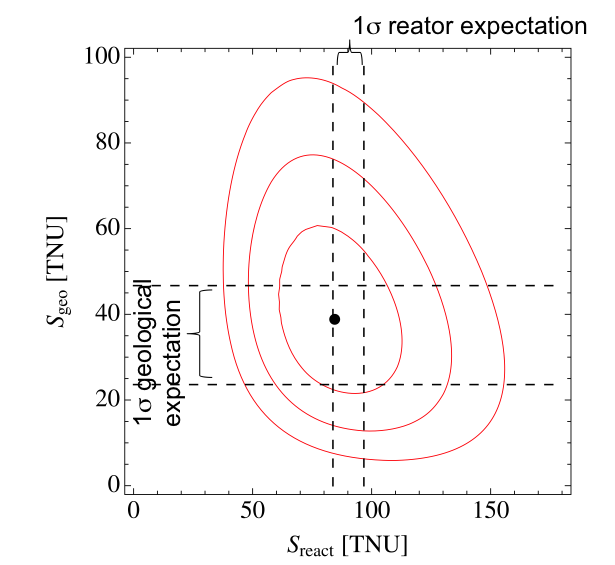}
\caption{Left: antineutrino spectrum (prompt positron scattering). Right: Unbinned maximum likelihood of geo vs reactor antineutrino events.} 
\label{fig:geo}
\end{center}
\end{figure}

Geo-neutrinos are anti-neutrinos produced in the radioactive chains of $^{238}$U, $^{ 232}$Th and by the decay of $^{40}$K, with a flux of the order of $\Phi_{\overline{\nu}}\sim 10^6$cm$^{-2}$s$^{-1}$.
These long lived elements are found in the Earth crust and mantel with unknown abundances. 
Measuring Geo-neutrinos flux at different locations can provide key information to the development of Earth models. 
Only geo-neutrinos from $^{238}$U and $^{ 232}$T chain elements can be measured by neutrino detectors, via Inverse Beta Decay with a threshold of 1.8MeV.  
The ratio of Th/U is supposed to be $\sim$3.9 from the analysis of meteorites with the same composition of the Earth. 
After the first observation in 2010, Borexino has now revised the result \cite{geo} with six times more statistics and a significantly refined modelling of the main background component: the antineutrinos from power reactors. 
Almost all other backgrounds are negligible thanks to the tagging of the events based on the coincidence between the prompt positron scattering and $\sim$250us delayed 2.2Mev gamma from the n capture on H. Events pairs are selected by energy cuts on both events and by their space-time correlation. 
The efficiency of the cuts is 0.84$\pm$0.01 from simulations. 
The enlarged dynamic fiducial volume (see above)  up to 25cm from the IV surface allows an exposure of (613$\pm$21) t $\cdot$ yr. 
We have selected 46 golden coincidences. An unbinned maximum likelihood fit of the prompt event energy spectrum returns $N_{react}=31.2^{+7}_{-6.1}$ (expected $33.3\pm2.4$) and $N_{geo}=14.3\pm4.4$. 
The last corresponds to a flux of $S_{geo}=38.8\pm12.0$ TNU (1TNU= 1$\nu$/10$^{32}$ protons /yr). 
The result is in good agreement with the Bulk Silicate Earth model predictions, however we are not yet at the level of discriminating among different model flavours.

\section{Phase II program}
\label{sec:phaseII}

The most important opportunity for Borexino phase II is the measurement of the neutrino flux from the fundamental pp reaction in the core of the Sun. 
This is made possible by the low $^{85}$Kr and $^{210}$Bi concentrations achieved with the purification campaigns. 
A dedicated effort has been made to understand the spectrum response in the $^{14}$C end-point region 
and its pile-up effects, disentangling it from the pp spectrum. 
The expected statistical error is below 10\% while the systematics are under study. 
The analysis is being finalised and the release is expected within 2014.
The second highest priority is the precision measurement of the pep flux, possibly with 10\% precision. 
At the same time we will attempt a measurement of the CNO fluxes, 
which are of fundamental astrophysical importance in particular as the they can help resolve the solar metallicity puzzle\cite{hax}. 
If this will not be possible, we foresee to sensibly improve the limits we have already posed and we will proceed to a further purification campaign to reduce the $^{210}$Bi contamination, which is the limiting factor of this analysis.
At the end of phase II we also expect to reduce the error on \ber flux at 3\% and to measure the seasonal variation effect upon several cycles and without the background constraints of phase I (sec. \ref{sec:annual}).
Finally the Geo-neutrino and the solar \bor fluxes can be measured with higher statistics, the error on the latter possibly being reduced below 10\%.

\section{Short distance Oscillation with boreXino (SOX)}
\label{sec:sox}

\begin{figure}[t]
\begin{center}
\includegraphics[width=0.47\linewidth]{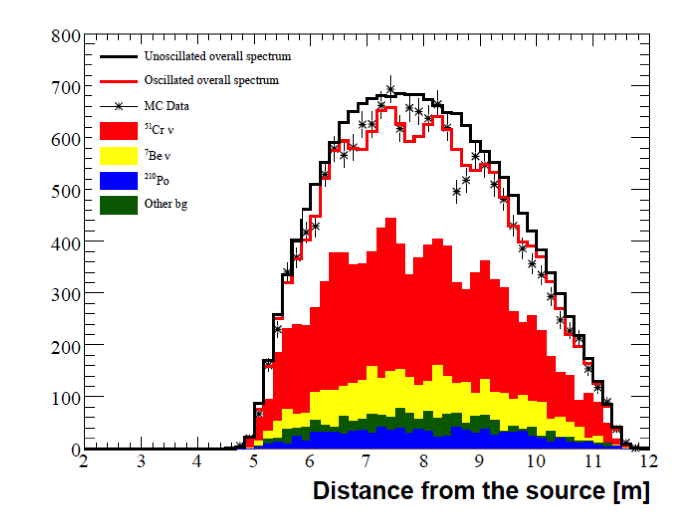}
\includegraphics[width=0.51\linewidth]{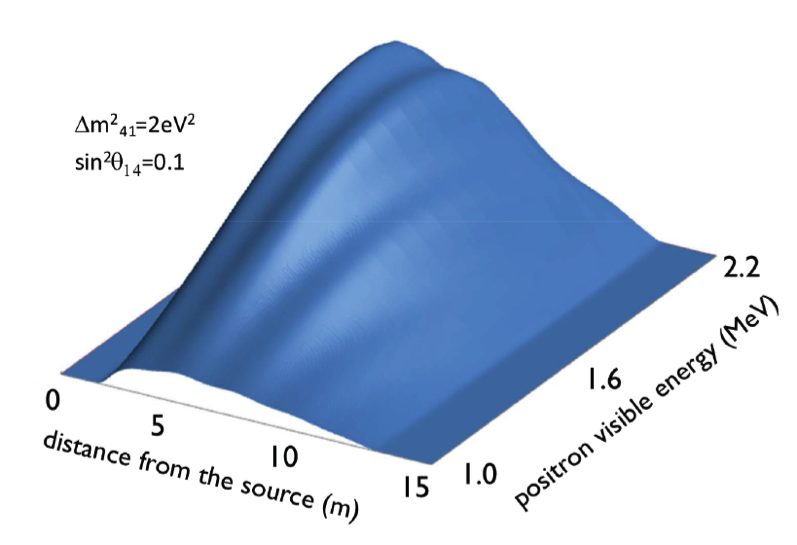}
\caption{
Oscillation wiggles in the SOX signals for a 10MCi \cro source (left) and 100kCi \cer source (right).}
\label{fig:sox_expected}
\end{center}
\end{figure}

\begin{figure}
\begin{center}
\includegraphics[width=0.53\linewidth]{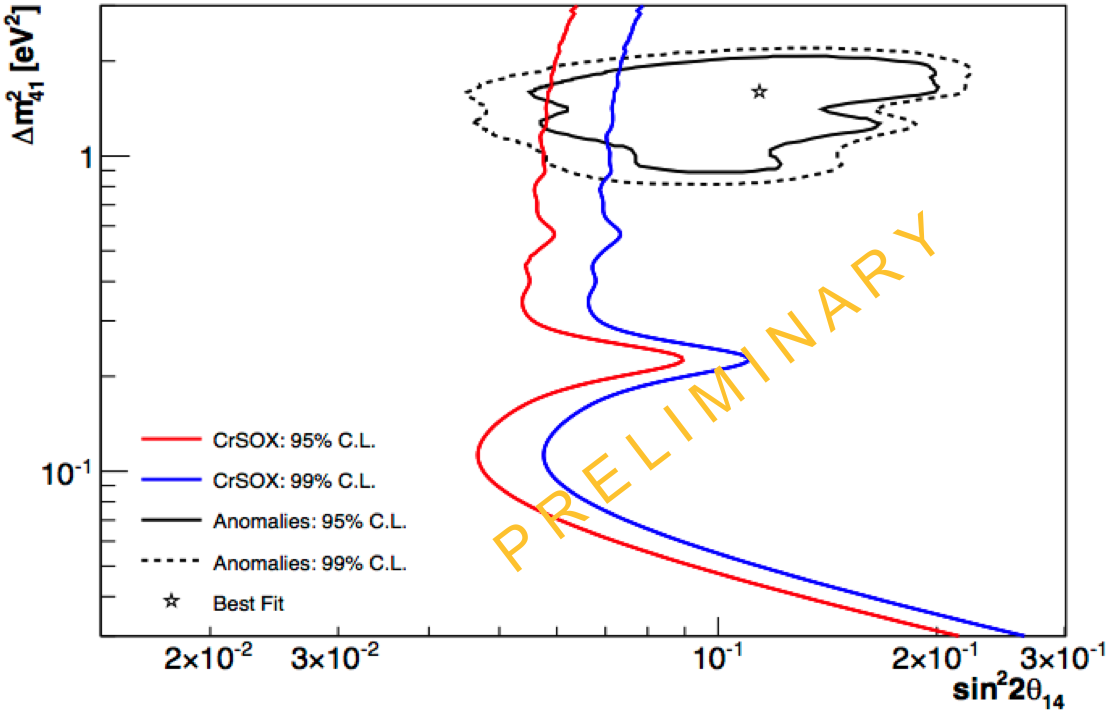}
\includegraphics[width=0.45\linewidth]{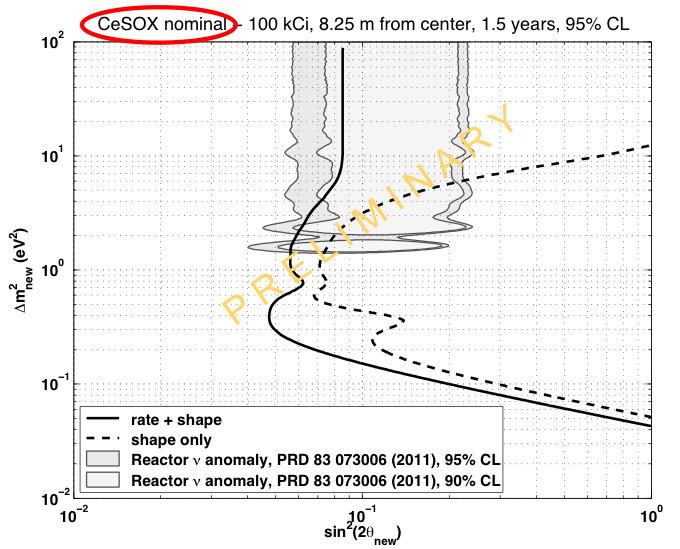}
\caption{SOX expected sensitivity. Left 10MCi \cro source compared to global analysis of all anomalies; Right 100kCi \cer source compared to reactor anomaly.}
\label{fig:sox_sensitivity}
\end{center}
\end{figure}

In the past years many different experimental indications have pointed toward the existence of sterile neutrinos. 
Although none of them is individually strong enough to make a claim, alltogether they justify a serious investigation. 
One of the most discussed evidence is the so-called reactor anomaly, 
which foresees an oscillation into the forth species with L/E$\sim$1m/MeV.
In Borexino, using neutrino source with energy of $\sim$1MeV, the oscillation length is significantly smaller then the detector size ($\sim$10m) and significantly larger then the detector resolution ($\sim$12-15cm). This allows to see oscillation wiggles in the position distribution of events. 
The location for such a source is the ~1m cubical pit present under the detector, which was excavated for this purpose before the detector's construction.
 This uninvasive deployment requires no work on the detector, it bears no risk of contamination and does not terminate the solar run of Borexino.
We foresee to deploy a \cro and a \cer source in 2015 and in 2016.
\cro is dichromatic neutrino emitter with energies of 430keV (10\%) and 750keV (90\%) and a relatively short decay time ($\tau\simeq40$d). 
The activity required is of the order of 10MCi and we plan to achieve this by re-activating the Chromium material used in Gallex and GNO experiments which has a 38\% $^{50}$Cr abundance.
Negotiations with reactor facilities in Oakridge (USA) and Mayak (Russia) are ongoing, taking also into account the need for quick transportation.
\cer instead is a $\beta$ emitter of antineutrinos with energies up to 3MeV and a more relaxed decay time of ~411d.
Thanks to the neutron tagging, which makes antineutrino detection essentially background free (see sec. \ref{sec:geo}), we can perform the measurement with a source activity of about 100-120kCi. 
Negotiation with the Mayak facility is ongoing, where a source of the required activity can be made out of spent nuclear fuel. 
Fig. \ref{fig:sox_expected} shows Montecarlo simulations of the expected signals for the two sources, while fig. \ref{fig:sox_sensitivity} shows the sensitivity of the two measurements in the oscillation parameters plane. Disappearance and wave effects will allow to clarify the matter, unambiguously proving or rejecting the hypothesis.
In particular, in case of the existence of a fourth sterile neutrino with parameters indicated by the reactor anomaly,
 SOX\cite{sox} will surely discover the effect and measure the parameters of oscillation.

\end{document}